\documentstyle[11pt,cs11,html,epsf]{article}

\def\solar{\ifmmode_{\mathord\odot}\else$_{\mathord\odot}$\fi}
\def\MM{$\cal{M}$\rm}
\def\MSUN{\MM\solar}

\def\etal{et al.\ }

\begin{document}

\title{Ages, Distances, and the Initial Mass Functions of Stellar Clusters}

\author{J. R. Stauffer}
\affil{Center for Astrophysics, 60 Garden St., Cambridge, MA 02138, USA}
\author{R. D. Jeffries}
\affil{Dept.\ of Physics, Keele University, Staffordshire, ST5 5BG, UK}
\author{E. L. Mart\'{\i}n}
\affil{Planetary Sciences Division, Caltech 105-21, Pasadena, CA 91125, USA}
\author{D. M. Terndrup}
\affil{Dept.\ of Astronomy, Ohio State University, Columbus, OH 43210, USA}

% Notice that some of these authors have alternate affiliations, which
% are identified by the \altaffilmark after each name.  The actual alternate
% affiliation information is typeset in footnotes at the bottom of the
% first page, and the text itself is specified in \altaffiltext commands.
% There is a separate \altaffiltext for each alternate affiliation
% indicated above.

% The nice thing about this method is that it saves space on the first page!

% BUT if you've used \altaffiltext and you think you'll be adding 
% *footnotes*, then you need to update the footnote counter!

%\setcounter{footnote}{3}

% The abstract is entered in a LaTeX "environment", designated with paired
% \begin{abstract} -- \end{abstract} commands.  Other environments are
% identified by the name in the curly braces.

\begin{abstract}

We provide a review of the current status of several topics on the
ages, distances, and mass functions of open clusters, with a
particular emphasis on illuminating the areas of uncertainty.
Hipparcos has obtained parallaxes for nearby open clusters that have
expected accuracies much better than has been previously achievable.
By using the lithium depletion boundary method and isochrone fitting
based on much improved new theoretical evolutionary models for low mass
stars, it is arguable that we will soon have have much better age
scales for clusters and star-forming regions.   With improved optical
and near-IR cameras, we are just now beginning to extend the mass
function of open clusters like the Pleiades into the regime below the
hydrogen burning mass limit.   Meanwhile, observations in star-forming
regions are in principle capable of identifying objects down to of
order 10 Jupiter masses.

\end{abstract}

% Keywords should be included, but they are not printed in the hardcopy.
% They will be used by the Editors to help organize poster papers by
% category though!

\keywords{stellar ages, stellar masses, open clusters}

% That's it for the front matter.  On to the main body of the paper.
% We'll only put in tutorial remarks at the beginning of each section
% so you can see entire sections together.

% 
% OK - to make things easier for the Editors, we're going to put
% all of our object aliases up front since we only have to declare
% them once in the paper.  Some people prefer to use NGC 7078 for
% M 15, but we like good old Messier, so that's what we'll index by.
% But we'll cross-reference it here so that people who do like NGC 7078
% won't have to remember that it's also M 15!
%
% Remember - we identify objects by putting an asterisk in front of the name!
%
\section{Introduction}

Rather than attempt to present an extremely abbreviated synopsis of all
of the material covered by the panel, we give here a summary of a few
of the presentations.  For each main topic, I (JRS) will give a brief
outline of the current paradigm and/or any well-known controversies.
Individual presentations by members of the panel will follow, with the
author of the contribution indicated by his initials at the end of
the section title.

\section{The Distances to Open Clusters}

The classical open cluster distance scale is based on determining an
astrometric distance to the Hyades, either via trigonometric parallax
or convergent point or some similar analysis, and then using
main-sequence fitting techniques to determine the distances of other
clusters relative to the Hyades (thus requiring a small correction for
the higher-than-solar metallicity of the Hyades).  This method
obviously requires an accurate knowledge of the cluster's reddening and
metallicity.   It further requires good photometry over the color range
where the cluster members lie on or near the ZAMS.  Finally, large
number statistics are useful in order that the single-star sequence be
well-defined.   The method assumes that the photometry for each cluster
is accurately calibrated, that the metallicities are accurately known
and the effects of metallicity correctly modeled, and that the
corrections for interstellar extinction are accurately derived from the
data.

For the nearer, richer open clusters, where photometry of large numbers
of stars is available (often in several different photometric systems),
the main sequence fitting distances are typically accurate to of order
5\% in the distance (0.1 mag in the distance modulus) unless there is
some significant error in the assumptions of the method.  Using the
available data, the commonly quoted distances to the Pleiades, Praesepe
and Coma clusters from main sequence fitting are of order 130 pc, 158
pc, and 80 pc, respectively.   Standard compilations of open cluster
distances include Becker \& Fenkart (1971), the Lynga catalog (Lynga
1987), and J.-C. Mermilliod's on-line ``Database for Stars in Open
Clusters'' (BDA) at 
{\tt http://cdsweb.u-strasbg.fr/online/mermio.html}\ .

\subsection{Comparison of Hipparcos Open Cluster Distances to MS
Fitting Distances (DMT)}

The high precision and large number of stars in the Hipparcos catalog
made it possible for the first time to derive trigonometric
distances to open clusters other than the Hyades (Robichon et 
al.\ 1999, van Leeuwen 1999), as well as an extremely precise 
(average) distance to the Hyades cluster
itself ($m - M = 3.34 \pm 0.01$, Perryman et al.\ 1998).  In several
cases, particularly for the Pleiades and Coma clusters, the derived
distances (118 pc and 90 pc, respectively for Pleiades and Coma) are
significantly different from distances inferred from the technique of
main-sequence fitting (MSF).  If the Hipparcos parallax to the Pleiades
were correct, then the main sequence of that cluster would be as much
as 0.3 mag fainter than expected from stellar evolution theory.

Prompted by this discrepancy, Pinsonneault et al. (1998) re-examined
the MSF distances to several open clusters with distances from
Hipparcos, and in particular discussed how the luminosity of the main
sequence depends on metallicity, helium abundance, and interstellar
extinction.  They concluded that none of these effects could produce
the anomalous main sequence luminosity, and suggested that there could
in some cases be systematic errors in the Hipparcos parallaxes on the
order of 1 milliarcsecond (mas), about 10 times larger than the
systematic errors in the parallaxes on a global scale.

Soderblom et al.\ (1998) examined the Hipparcos data base, selecting
young main-sequence stars with parallax errors small enough that
systematic errors at the 1 mas level would be unimportant.  If nature
were able to produce subluminous stars in the Pleiades, they reasoned,
then there should be similar objects in the field.  They were unable to
find any, further suggesting that the Pleiades distance from Hipparcos
was incorrect.
 
In the design of the Hipparcos experiment, it was known that a
potential problem in determining the distances to open clusters is the
possibility of correlated errors on small angular scales (e.g.,
Lindegren 1988).  Furthermore, the typical density of stars in open
clusters is significantly higher than the average across the sky.
(Both these effects were included in the Hipparcos analysis of the
cluster distances.)  Narayanan \& Gould (1999a) proposed a new method
of examining the Hipparcos parallaxes in open clusters.  Applying this
to the Pleiades and Hyades (Narayanan \& Gould 1999b), they showed that
the Hipparcos parallaxes toward these open clusters are spatially
correlated over angular scales of $2^\circ - 3^\circ$, with an
amplitude of up to 2 mas.  This correlation is stronger than expected
based on the analysis of the Hipparcos catalog.  Using a distance
method based on the Hipparcos proper motions which should not be biased
by these spatial correlations, Narayanan \& Gould derived a distance to
the Pleiades essentially equal to the main-sequence fitting distance
but significantly different from that derived from the Hipparcos
parallaxes.   For the Hyades, by chance, the structure of the spatial
correlations is such that the errors average out and the distance
derived from the Hipparcos parallaxes agrees with that derived from
main-sequence fitting.

The issue is not settled, however, because those most closely connected
to the Hipparcos parallax measurements maintain that their distances
are correct, and that there is an astrophysical explanation for the
discrepant Pleiades and Coma cluster distances (see the talk in these
proceedings by M. Grenon).   Fortunately, there is reason to believe that
a definitive resolution can now be expected from new observations because
NASA has approved construction of a new astrometric mission called FAME,
which should obtain parallaxes for many more stars and to significantly
higher accuracy than Hipparcos -- and provide the answers within the next
decade based on an expected 2004 launch 
(see {\tt http://aa.usno.navy.mil/fame/}).

\section{Open Cluster Ages}

The traditional open cluster age scale is based on fitting photometry
of stars near the upper main sequence turnoff of a given cluster to
theoretical evolutionary isochrones.   The most widely referenced
compendium of open cluster ages derived in that manner is probably
Mermilliod (1981).  In that paper, the ages of the Alpha Persei,
Pleiades, and Hyades open clusters are listed as 51 Myr, 78 Myr, and 660
Myr, respectively.  While other age indicators confirm the relative
ordering of cluster ages given by Mermilliod, the absolute age scale
has been a subject of some controversy.  Most particularly, by
adjusting the amount of ``convective core overshoot" for stars near the
MS turnoff, it is possible to derive cluster ages up to a factor of two
older than those given in Mermilliod (1981):  see, for example, 
Mazzei \& Pigatto (1989) or Meynet \etal (1993).

An alternate way to estimate cluster ages is by determining the
location of the pre-main sequence (PMS) turn-on point or the
displacement of the PMS locus above the ZAMS.  Because of fairly
obvious realities, this method has mostly been applied to very young
clusters.  However, a few -- perhaps foolhardy -- authors have
attempted to apply the method to ``oldish" clusters like the Pleiades
or Alpha Persei (Stauffer 1984; Stauffer \etal 1989).   With the
development of more accurate theoretical evolutionary models that
extend to lower masses, it is now becoming more feasible to extend this
type of age-dating to ``oldish" open clusters, and two discussions of
this topic are included below.

A new way to estimate the age of open clusters has recently been
proposed by Basri, Marcy, \& Graham (1996=BMG), Bildsten \etal (1997),
Ushomirsky \etal (1998), Ventura \etal (1998) and others.  The idea
behind this method is that for stars near the substellar mass limit,
the age at which stars become hot enough in their cores to burn lithium
is a sensitive function of mass.  Furthermore, it is argued that the
physics required to predict the location of the ``lithium depletion
boundary" (LDB) as a function of age is very well understood and not
subject to significant uncertainty (Bildsten \etal 1997).  In the
mass range of interest, stars are fully convective, and the core
lithium abundance will be directly reflected in the surface lithium
abundance - and the latter can be determined by use of the
6708\AA\ \ion{Li}{1} doublet.

BMG and Rebolo \etal (1996) were the first to successfully apply this
test by detecting lithium in three substellar objects in the Pleiades.
However, it was later determined (Basri \& Mart\'{\i}n 1999a) that the
exact location of the lithium boundary (and hence the Pleiades age)
was uncertain because the brightest of the three objects (PPL15) is a
nearly equal mass binary.  This problem was resolved by Stauffer,
Schultz, \& Kirkpatrick (1998=SSK), who obtained spectra of an
additional 10 faint Pleiades members, five of which still retain their
lithium and by Mart\'{\i}n \etal (1998) who obtained spectra of an
additional two Pleiades members, one of which had detected lithium.  By
providing measurements of a large number of stars near the lithium
boundary, SSK claimed to determine the absolute magnitude of the
lithium depletion boundary to an accuracy of about 0.1 mag,
corresponding to an age uncertainty of about 8 Myr.  The age derived
for the Pleiades by SSK was 125 Myr.

More recently, Basri \& Mart\'{\i}n (1999b) and Stauffer \etal (1999)
have determined an LDB age for the Alpha Persei cluster (based,
respectively, on one and five members with detected lithium), and
Barrado y Navascu\'{e}s, Stauffer \& Patten (1999) have derived an LDB
age for the IC2391 open cluster.  The ages for all three clusters
derived in this manner are systematically older (by of order 50\%) than
the ages quoted in Mermilliod (1981).  If these LDB ages are as
accurate and precise as believed, it should be possible to define a new
open cluster age scale to replace Mermilliod (1981).  Two of the
panelists however urge caution at this point.

\subsection{Limitations of the Lithium Depletion Boundary Method (RDJ)}

Whilst the Lithium Depletion Boundary Method is potentially a very
precise way of determining the age of a cluster, it is worth
considering in some detail to what extent the age error budget is
influenced by various sources of uncertainty. These can be placed into
two categories; random and systematic.

\begin{figure}
\plottwo{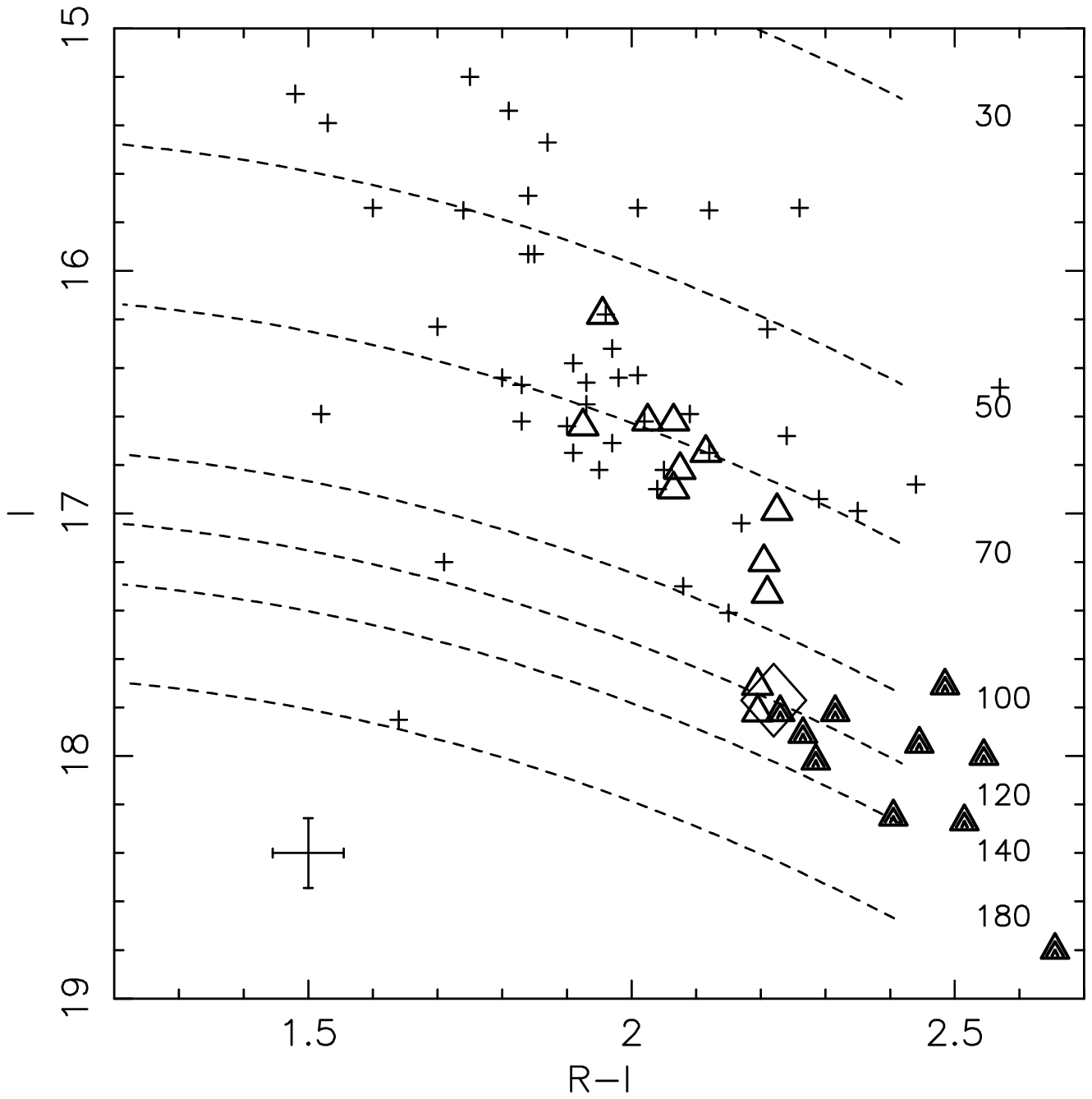}{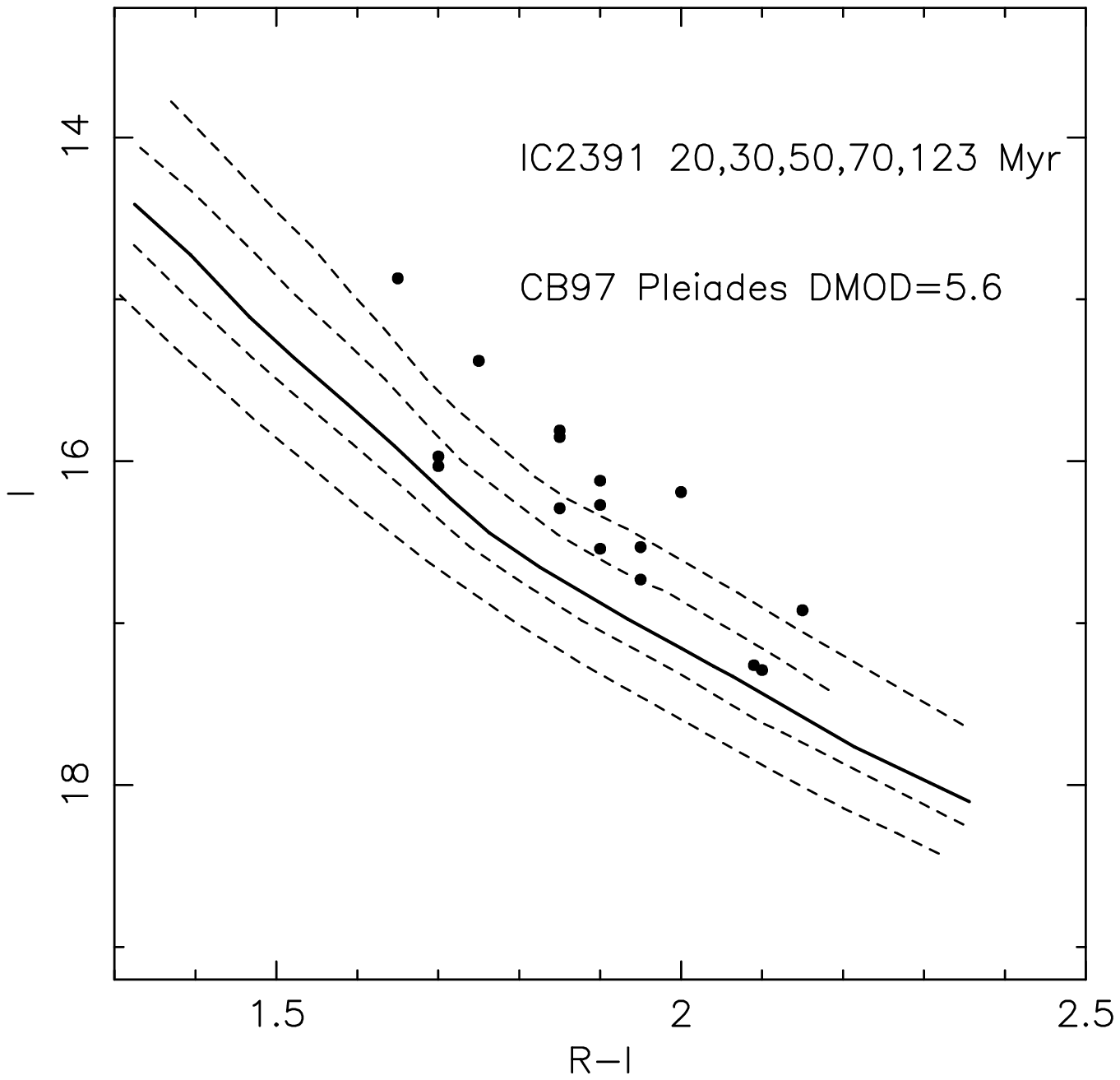}
\caption{(a) CMD for the Pleiades showing Li detections (filled
triangles), non-detections (open triangles) and members without Li
measurements (crosses). The diamond indicates the assumed location of the
LDB and its error. The error bar indicates the effects of other random
errors (distance modulus, photometric calibration etc.). Dashed lines
are isochrones of 99\% Li depletion from Chabrier \& Baraffe (1997),
labelled in Myr. (b) A CMD for IC 2391. Isochrones are plotted for a
number of ages including the LDB age (thick line). Isochrones are
derived from Chabrier \& Baraffe models, with an $R-I$-$T_{\rm eff}$
relation calibrated on a Pleiades LDB isochrone of 123\,Myr.}
\label{rdjplot1}
\end{figure}

My baseline assumption is that the age is found by locating the LDB in
a colour-magnitude diagram (CMD), using the colour of the LDB and an
empirical bolometric correction (BC) to find $L_{\rm bol}$ at the LDB
and then using a model for Li depletion in cool stars to translate this
into an age (see Fig.~\ref{rdjplot1}a). Random errors can be ascribed
to uncertain placement of the LDB in the CMD, uncertain photometric
calibrations and uncertainties in the distance and reddening of the
cluster in question. Systematics arise from the chosen BC-colour
relation, whether one assumes the LDB defines the point at which say
90\% or 99\% of Li has been depleted and the choice of evolutionary
model which defines the $L_{\rm bol}$-age relation at the LDB. The
first two of these systematics will alter the ages of all clusters in
qualitatively the same way, whereas the latter systematic may cause
individual  clusters to be older or younger, as the $L_{\rm bol}$-age
relations cross for differing evolution models. There may also be
additional systematics due to common assumptions made by {\em all} the
current generation of models.

I have investigated these uncertainties, focussing on clusters with an
LDB defined in the $I$ vs $R-I$ CMD, and using the theoretical models
of Burrows et al.\ (1997), Chabrier \& Baraffe (1997) and D'Antona \&
Mazzitelli (1997). Note that model colours and magnitudes from the
former two sets of models have not been used as their uncertainties are
unquantified. I find that for clusters with data quality similar to
that available for the Pleiades, random errors in the age rise from
$\pm 9\%$ at $\sim 25$\,Myr, to $\pm 16\%$ at 200\,Myr. These errors are
dominated by uncertain placement of the LDB in the CMD ($\pm 0.15$ mag
in $I$ and $\pm 0.05$ in $R - I$) along with distance modulus errors and
uncertainties in the $I$ photometric calibration (both around $\pm0.1$
mag). Obtaining Li data for more points around the LDB, distinguishing
binaries from single stars and tightening up the photometric
calibration could reduce these errors considerably. The systematic
errors are more constant, ranging between $7\%$ and $11\%$ over a
similar age range.  The systematics are due mainly to the choice of
model and whether there is any age dependence in the BC-colour
relation. The influence of whether Li is depleted by 90\% or 99\% at
the LDB (say for Li 6708\AA\ EWs $<0.3$\AA) only alters ages by
$\pm 3$\%.

In Table~\ref{ldbages}, I present the results of my analysis for the
Pleiades, Alpha Per and IC 2391 clusters using Li spectra, $I$ and
$R-I$ data from the literature and averaging over the three evolution
models. These ages are similar to those previously estimated (Stauffer
et al.\ 1998, 1999; Barrado y Navascu\'{e}s et al.\ 1999), but the
estimated errors are larger by a more than a factor of two. This does
not alter the conclusion that LDB ages are older than turn-off ages
with no core overshoot, but certainly obscures whether core overshoot
might be mass dependent (see \S 3.2).

I also have evidence that there may be additional systematics in the
LDB method that are not adequately reflected by simply choosing a
variety of models. If I assume the Pleiades age is given by its LDB age
for a given set of evolutionary tracks, then I can force an isochrone
in the CMD to match the available cluster photometry by choosing a
particular form of the $R-I$-$T_{\rm eff}$ relation. The only other
parameters involved here are cluster distance and reddening and the
empirical BCs -- none of which are uncertain enough to affect my
conclusions. If I further assume that this $R-I$-$T_{\rm eff}$ relation
applies to younger stars, in IC 2391 for example, then the isochrone
defined by their LDB age should coincide with the observed cluster CMD.
If it does not then there is either something wrong with the evolution
models or the assumption that a single colour-$T_{\rm eff}$ relation
holds over a range of gravity is flawed. Figure~\ref{rdjplot1}b shows
the result of doing this, assuming a Pleiades distance modulus of 5.6.
There is clearly a discrepancy. Using the Hipparcos distance modulus of
5.3 makes the Pleiades LDB age older by $\sim 25$\,Myr and merely
increases the discrepancy.  I conclude that there may yet be a
systematic problem in the evolutionary models (all of them) that could
be solved if IC 2391 were younger than suggested by its LDB age. An
age-dependent colour-$T_{\rm eff}$ relation seems less likely as it is
not predicted in currently published models and spectroscopic indices
in the $R$ and $I$ bands give reliable estimates of $R-I$ colours in
all these clusters. This suggests that any gravity dependence of
spectral features is heavily subordinate to the dominant $T_{\rm eff}$
dependence.

\begin{table}
\caption{LDB ages for open clusters}\label{ldbages}
%\begin{center}
\begin{tabular}{lccc}
           & Age   & Random Error & Systematic Error  \\
Cluster    & (Myr) &   (Myr)      &  (Myr) \\ \tableline
Pleiades   & 122   &  $\pm 18$    & $\pm 11$               \\
Alpha Per  &  85   &  $\pm 11$    & $\pm  8$               \\
IC 2391    &  48   &  $\pm  5$    & $\pm  3$               \\ \tableline
\end{tabular}
%\end{center}
\end{table}

\subsection{Age Estimates from Isochrone Fitting (DMT)}

\begin{figure}\label{aper-cmd}
\epsscale{0.6}
\plotone{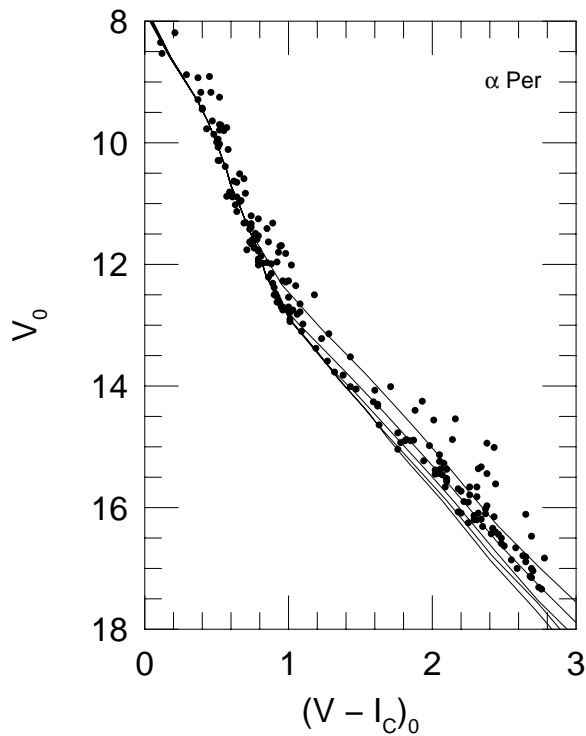}
\caption{Color-magnitude diagram for the Alpha Persei cluster compared
to isochrones of various ages.  The photometry has been dereddened and
the isochrones are for a distance modulus of $m - M = 6.25$.  Ages are
40, 60, 80, 10, and 120 Myr; the youngest ages are brightest.}
\end{figure}

There are new efforts underway to compare the MSF and Hipparcos
distances for more clusters.  For example, Pinsonneault et al.\ (2000)
have expanded their MSF technique, which employs the Yale YREC
isochrones, to work over a wider range of luminosity on the main
sequence than previously, by recomputing color-effective temperature
transformations to reproduce the morphology of the Pleiades main
sequence.  For clusters younger than the age of the Hyades, the
technique can be used to estimate the age of the cluster, since the
lowest-mass stars are still descending towards the main sequence.  This
technique is illustrated in Figure \ref{aper-cmd}, which shows a
color-magnitude diagram for Alpha Per compared to isochrones of various
ages.  An age of $60 \pm 6$ Myr results.  This method also generates a
metallicity estimate for each cluster by comparing the distance derived
using $B - V$ photometry to that obtained in $V - I$; since the
luminosity depends on metallicity differently in the two colors, one
derives distances which disagree unless the metallicity is correct.

Pinsonneault et al.\ (2000) derive distances and photometric
metallicities for Alpha Per, the Pleiades, NGC 2516, NGC 6475, and
NGC 6633; for other clusters they adopt the high-resolution
spectroscopic abundances of Boesgaard \& Friel (1992) and the distances
computed in Pinsonneault et al.  (1998).  There are some systems (the
Pleiades and Coma Ber) which still have large differences between the
MSF and Hipparcos distances, while there are others (NGC 6475 and
Alpha Per) which may possibly be in conflict depending on the
treatment of the Hipparcos errors.  There is no obvious pattern to the
deviations with age or metal abundance, and furthermore the differences
are not consistent with a simple scale shift to systematically shorter
or longer distances.
 
\begin{figure}
\epsscale{0.9}
\plotone{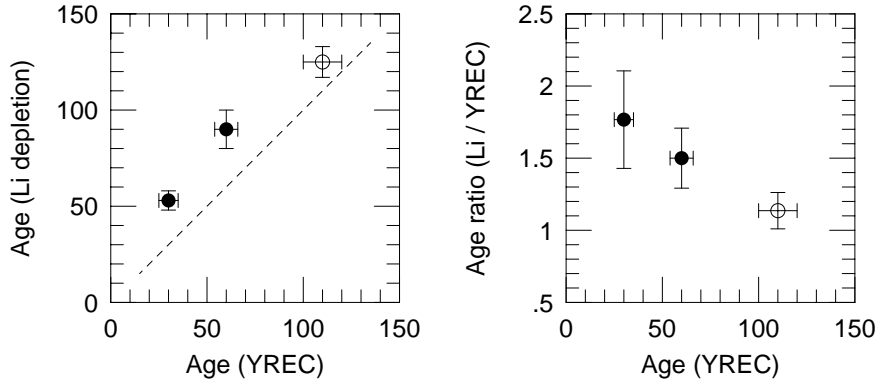}\label{age-compare}
\caption{(a) Direct comparison of ages derived from the LDB method
to ages derived from isochrone fitting to the YREC theoretical models.
(b) Ratio of the LDB age to the YREC age, as a function of age.}
\end{figure}

In most cases, the ages derived in Pinsonneault et al.  (2000) are
somewhat higher than had been found from the main-sequence turnoff (as
in Mermilliod 1981), but are still rather younger than the ages
inferred from the LDB method.  The two ages are compared in Figure
\ref{age-compare} for three open clusters (from youngest to oldest
being Alpha Per, IC 2391, and the Pleiades).  The upper panel compares
the ages directly, with the dashed line indicating equality.  The ratio
of the two ages is shown in the lower panel; there is probably an
age-dependent offset in the two scales.

\section{Cluster and Star-Forming Region Mass Functions}

It is only within the past few years that open cluster mass functions
that extend to low masses ($\sim$0.1-0.2 \MSUN) and that are based on
proper motions have started to become available.   This situation has
resulted from a lack of the appropriate raw materials (deep, 1st and 2nd
epoch imaging data) and a lack of sufficiently good and automated
measuring machines (cf. Hambly 1998).  Much of the literature for open
clusters in fact has not been directed at determining a mass function,
but instead has concentrated on the more directly determined luminosity
function.  At least up until 1980, and possibly up until much later,
the most cited primary reference for open cluster luminosity functions
was the work by van den Bergh \& Sher (1960).  Those authors estimated
luminosity functions for a large number of open clusters using
photographic plates and estimated cluster-member data derived from
counting the number of stars in annuli centered on the apparent cluster
center.  van den Bergh and Sher's conclusion from these data was that
most open clusters had falling or flat luminosity functions faintward
of M$_{B}$ $\sim$ 6, and hence that open clusters are quite deficient
in low mass stars compared to the field.  This result was later used as
one piece of evidence in favor of bimodal star formation -- in
particular in favor of the idea that low mass stars are formed in loose
associations (like Taurus-Auriga) and high mass stars are formed in
dense clusters.  However, the observational result in fact was almost
certainly wrong because the outer annuli used to derive the ``field"
star density was chosen at too small a radius (which preferentially
affects the derived number of faint cluster members due to mass
segregation).

The situation in star-forming regions was not much better until quite
recently, again due to a combination of lack of the right equipment
(mostly sensitive, low-noise IR detectors) and lack of the
right apparatus to utilize those observations (in this case, 
theoretical evolutionary tracks and ways to compare them to
observations for very young ages).  Much progress has been made in the
past few years on both fronts, with the results that it is possible to
begin to have confidence in the mass functions being derived for
regions like rho~Oph and IC~348.  It is particularly likely that the
derived mass functions will be close to reality when spectra have been
obtained for all or most of the low mass objects so that the estimated
effective temperatures are determined better and non-members can be
excluded (see especially recent papers by K.\ Luhman on this topic).

\subsection{Open Cluster Mass Functions (RDJ)}

The problem of determining cluster mass functions (MFs) can be broken
down into three parts: surveying for cluster members, converting the
observed luminosity function into a MF and then correcting this MF for
any mass-dependent spatial dispersion of stars (mass segregation).

A survey for cluster members might ideally use proper motions,
photometry, radial velocities, spectral types, the presence of Li,
X-ray emission or some subset of these to form an opinion on individual
membership. It is clearly of vital importance to understand both the
{\em completeness} (to include members) and {\em specificity} (to
exclude non-members) of any membership tests. These need not be known
on an individual basis, but must at least be known statistically as a
function of luminosity (e.g., Hambly et al.\ 1999).

The conversion from luminosity functions to a MF is not trivial,
relying on the age and distance of the clusters (with age becoming more
important in younger clusters), a choice of stellar evolution models
and a knowledge of the fraction of stars that are unresolved binary
systems (with consequently larger luminosities).

The final step of correcting for the effects of mass segregation is
often forgotten. For instance, the Pleiades should be almost
dynamically relaxed and equipartition will have ensured that lower mass
stars have a larger core radius ($r_{c} \propto M^{-1/2}$). This effect
has been detected by Pinfield et al.\ (1998) and could have a large
effect on the low mass end of the mass function.  Unfortunately, the
relaxation time also becomes longer at lower masses and this combined
with the relatively low number of brown dwarfs uncovered in the
Pleiades makes normalization of the brown dwarf MF in the Pleiades
uncertain by a factor of $\sim 2$ relative to the higher mass stars
(Hodgkin \& Jameson 2000). Finally, there is of course the possibility
of preferential ejection of lower mass cluster members that stray
beyond the tidal radius. There is no alternative to a theoretical
correction for this effect if one wishes to estimate the {\em initial}
MF from a present-day MF. Clearly these dynamical effects have
considerably less influence in younger clusters and associations.

The state of the art MF for open clusters is that derived in the
Pleiades (Bouvier et al. 1998, Hambly et al.\ 1999, Hodgkin \& Jameson
2000). This now extends from the A stars down to brown dwarfs at
$\simeq 0.04M_{\odot}$. Whilst there is still disagreement on the exact
representation of the MF, a number of clear points have emerged. (i)
The MF cannot be represented as a single power law, but a log-normal
MF, similar to that proposed by Miller \& Scalo (1979) certainly does a
reasonable job down to 0.08$M_{\odot}$. (ii) A power law fit to the
brown dwarf MF still has uncertainties of 50\% or so in the power law
index because of the uncertainties in membership, small number
statistics and uncertain dynamical effects. However, if the cluster MF
is similar to the field MF, it is clear that brown dwarfs
($0.04 < M < 0.08 M_{\odot}$) contribute negligibly to the Galactic
disk mass.  (iii) There is no evidence for a steep turnover or
truncation of the MF down to 0.04$M_{\odot}$.

\subsection{Comments on the Mass Function of the Pleiades (ELM)}

My collaborators and I have obtained new IR photometry and optical
spectroscopy of all of the very low mass and brown dwarf candidate
members of the Pleiades proposed in Bouvier \etal (1998).  We have used
those new data (Mart\'{\i}n \etal 2000) 
to attempt to determine which of these stars are real
members of the cluster, and which are instead field star contaminants.
The criteria used include spectral type and luminosity class, presence
and strength of H$\alpha$\ emission, and location in color-magnitude
diagrams.  In addition, we consider other published information such as
the presence of lithium in absorption and radial velocity.

Table~\ref{cfhdat}  provides a summary of the membership information
for these stars.   The bottom line is that the success rate for the
Bouvier \etal candidate list is in the range 63$\%$ to 78$\%$,
depending on the status of the YES? stars (for which the results are
currently not definitive).   Bouvier \etal had predicted a probable
success rate of 75$\%$ based on a simple field star luminosity function
and survey volume argument.  The new data are consistent with the
original estimate, with a possibility of somewhat more field star
contamination than expected.   Therefore, the Bouvier \etal estimate of
the Pleiades IMF down to 0.04 M$_{\odot}$\ is supported by the new
membership study, indicating a slightly rising mass function below the
hydrogen burning mass limit.  This mass function is slightly less steep
than the mass function derived by Mart\'{\i}n \etal (1997), based on
the IAC Pleiades surveys which concentrated on fields somewhat closer
to the cluster center than the CFHT fields.  This could be a hint that
the brown dwarfs are more concentrated to the cluster center than the
VLM stars.

\begin{table}
\caption{Membership Assessment in the Pleiades}\label{cfhdat}
\begin{center}\scriptsize
\begin{tabular}{lcccccccccc}
     & Other &     &      &   &    &     &
Spec. & $I$ vs.   & NIC   &   \\ 
Name & Name  & Dwarf? & V$_{\rm rad}$ & pm & Li & H$\alpha$ &
type & $I-K$ & Seq  & Member?  \\ \tableline 
CFHT  1       &          & Yes &     &     &     & Yes  & Yes &  Yes &     & Yes \\
CFHT  2       &          & Yes &     &     &     & Yes  & Yes &  Yes &     & Yes \\
CFHT  3\tablenotemark{*}       & HHJ 22   & Yes &     & Yes &     &      &     &  Yes &     & Yes  \\
CFHT  4       &          &     &     &     &     &      &     &  Yes &     & Yes? \\
CFHT  5       &          & Yes &     &     &     & Yes  & Yes &  Yes &     & Yes \\
CFHT  6       &          & Yes &     &     &     & No   & Yes &  Yes &     & Yes? \\
CFHT  7       &          & Yes &     &     &     & No   & Yes &  Yes &     & Yes? \\
CFHT  8       &          & Yes &     &     &     & Yes  & Yes &  Yes &     & Yes \\
CFHT  9       &          & Yes &     &     & No  & Yes  &     &  Yes &     & Yes \\
CFHT 10       &          & Yes &     &     & No  & Yes  &     &  Yes &     & Yes \\
CFHT 11       & Roque 16 & Yes &     &     & Yes & Yes  &     &  Yes & Yes & Yes \\
CFHT 12       &          & Yes &     &     & Yes & Yes  &     &  Yes & Yes & Yes \\
CFHT 13       & Teide 2  & Yes & Yes &     & Yes & Yes  & Yes &  Yes & Yes & Yes \\
CFHT 14       &          & Yes &     &     & No  & No   &     &      &     & No  \\
CFHT 15       &          & Yes &     &     & Yes & Yes  &     &  Yes & Yes & Yes \\
CFHT 16       &          & Yes &     &     &     &      & No  &  Yes?  & Yes?  & Yes?  \\
CFHT 17       &          & Yes &     &     &     & Yes  & Yes &  Yes & Yes & Yes \\
CFHT 18       &          & Yes &     &     & No  & Yes  & Yes &  Yes & No  & No \\
CFHT 19       &          & Yes &     &     &     &      & Yes &  No  & Yes & No \\
CFHT 20       &          & Yes &     &     &     & No   & Yes &  No  & No  & No \\
CFHT 21       & Calar 3  & Yes & Yes &     & Yes & Yes  & Yes &  Yes & Yes & Yes \\
CFHT 22       &          & Yes &     &     &     &      & Yes &  No  & No  & No \\
CFHT 23       &          & Yes &     &     &     &      &     &  Yes & Yes & Yes \\
CFHT 24       & Roque  7 & Yes &     &     &     &      & Yes &      & Yes & Yes \\
CFHT 24.1     & PIZ 1    & Yes &     &     &     &      & Yes &  Yes & Yes & Yes \\
CFHT 25       &          & Yes &     &     &     & No   & Yes &  Yes & Yes & Yes \\
CFHT 26       &          & Yes &     &     &     & No   & No  &      &     & No  \\
KPNO 3\tablenotemark{*}        & HHJ 20   & Yes &     & Yes &     &      &     &  Yes &     & Yes  \\
KPNO 4\tablenotemark{*}        & HHJ 28   & Yes &     & Yes &     &      &     &  Yes &     & Yes  \\
KPNO 5        &          &     &     &     &     &      &     &      &     & Yes?  \\ 
\end{tabular}
\end{center}
\tablenotetext{*}{Dwarf status inferred from proper motion.}
\end{table}

%
%

% That's all, folks.
%
% The technique of segregating major semantic components of the document
% within "environments" is a very good one, but you as an author have to
% come up with a way of making sure each \begin{whatzit} has a corresponding
% \end{whatzit}.  If you miss one, LaTeX will probably complain a great
% deal during the composition of the document.  Occasionally, you get away
% with it right up to the \end{document}, in which case, you will see
% "\begin{whatzit} ended by \end{document}".

\end{document}